\documentclass[11pt, oneside]{article}   	% use "amsart" instead of "article" for AMSLaTeX format
\usepackage[margin=1.1in]{geometry}                		% See geometry.pdf to learn the layout options. There are lots.
\usepackage{graphicx}				% Use pdf, png, jpg, or eps§ with pdflatex; use eps in DVI mode
\usepackage{amssymb}

\title{Likelihood-ratio inference on differences in quantiles}
\author{Evan Miller\thanks{Currently at Anthropic, \texttt{evanmiller@anthropic.com}} \\ Eppo \\ \texttt{evan@geteppo.com}}
\begin{document}
\maketitle
%\section{}
%\subsection{}

Quantiles can represent key operational and business metrics, but the computational challenges associated with inference has hampered their adoption in online experimentation. One-sample confidence intervals are trivial to construct; however, two-sample inference has traditionally required bootstrapping or a density estimator. This paper presents a new two-sample difference-in-quantile hypothesis test and confidence interval based on a likelihood-ratio test statistic. A conservative version of the test does not involve a density estimator; a second version of the test, which uses a density estimator, yields confidence intervals very close to the nominal coverage level. It can be computed using only four order statistics from each sample.

\section{Background}

Notation follows \cite{spotifybootstrap}. Let $0<q<1$ represent the population quantile of interest, with $\tau_{q,c}$ and $\tau_{q,t}$ representing the true quantile values in control and treatment groups. Suppose that ordered samples $\mathbf{\tilde{y}}_c=(y_{c,(1)},\ldots,y_{c,(N_c)})$ and $\mathbf{\tilde{y}}_t=(y_{t,(1)},\ldots,y_{t,(N_t)})$ are drawn from control and treatment.

The traditional Price-Bonnet \cite{price2002} estimator forms a confidence interval on the difference in quantiles as

\[
C(\alpha)=\hat{\tau}_{q,t}-\hat{\tau}_{q,c}\pm z_{\alpha/2}\sqrt{{\rm Var}(\hat{\tau}_{q,t})+{\rm Var}(\hat{\tau}_{q,c}})
\]

Donner and Zou \cite{zou2008} noted that this method fails to account for asymmetry in the underlying sampling distributions. They propose instead 

\[
C^+(\alpha)=\hat{\tau}_{q,t}-\hat{\tau}_{q,c}+\sqrt{(u_t(\alpha)-\hat{\tau}_{q,t})^2+(\hat{\tau}_{q,c}-l_c(\alpha))^2}
\]
\[
C^-(\alpha)=\hat{\tau}_{q,t}-\hat{\tau}_{q,c}-\sqrt{(\hat{\tau}_{q,t}-l_t(\alpha))^2+(u_c(\alpha)-\hat{\tau}_{q,c})^2}
\]

where $l_x(\alpha)$ and $u_x(\alpha)$ represent the lower and upper bounds of the one-sample $100(1-\alpha)\%$ confidence interval for group $x$. Simulations indicate that the Donner-Zou method yields anti-conservative CIs. The method proposed below is similar in spirit to Donner-Zou, but operates in index space rather than value space. It does not involve the quantile estimates themselves, and simulations show that it has better coverage properties than Donner-Zou.

\section{Setup}

Suppose an unknown distribution $F$ has true quantile value $\tau_q$. If a sample of size $N$ is drawn from $F$, then the probability that the sample contains exactly $i$ observations less than $\tau_q$ is given by the binomial distribution. Then given an ordered sample $\mathbf{\tilde{y}}=(y_{(1)},\ldots,y_{(N)})$, a likelihood function may be constructed as

\[
L(\tau_q;y_{(i)}<\tau_q<y_{(i+1)})=h(i|q,N)={N \choose i}q^i(1-q)^{N-i}
\]

Note that this likelihood function is flat between points in the ordered sample, and so a unique maximum does not exist.

Next consider two samples $\mathbf{\tilde{y}}_c$ and $\mathbf{\tilde{y}}_t$ and a constraint (hypothesis) of the form $\tau_{q,t}=\tau_{q,c}+d$. We can test this hypothesis with a likelihood-ratio test, maximizing the joint (two-sample) likelihood with and without the constraint applied.

\begin{figure}[h]
\centering
\includegraphics[width=\textwidth]{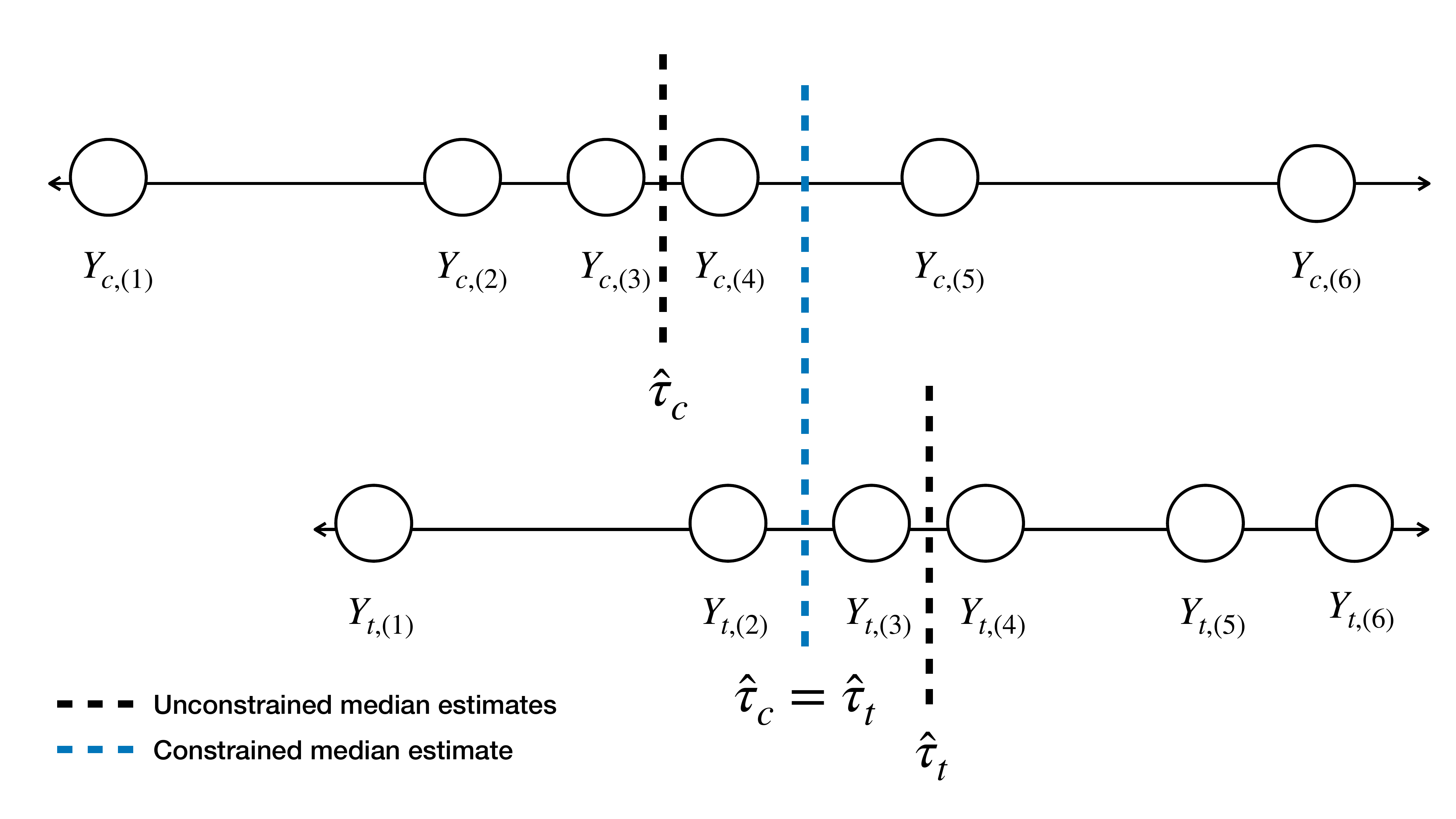}
\caption{Constrained and unconstrained MLE estimates of the true median value}
\end{figure}

The unconstrained likelihoods will be maximized in a region near the sample quantiles. If $q(N+1)$ is an integer, the one-sample likelihood for $\tau_{q}$ will be maximized over the open interval $\hat{\tau}_q\in (y_{(qN)},y_{(q(N+2))})$. If $q(N+1)$ is not an integer, the one-sample likelihood will be maximized over the open interval $\hat{\tau_q}\in (y_{(\lfloor q(N+1) \rfloor)}, y_{(\lceil q(N+1) \rceil)})$. In either case, the maximizing likelihood can be written $h(\lfloor q(N+1) \rfloor|q,N)$, and the maximized unconstrained joint likelihood is given by

\[
L(\hat{\tau}_c,\hat{\tau}_t)=h(\lfloor q(N_c+1) \rfloor|q,N_c)\times h(\lfloor q(N_t+1) \rfloor|q,N_t)
\]

The constrained problem will maximize the joint likelihood

\[
L(\tau_{q,c},\tau_{q,t}; y_{c,(i)}<\tau_{q,c}<y_{c,(i+1)}, y_{t,(j)}<\tau_{q,t}<y_{t,(j+1)})=h(i|q,N_c)\times h(j|q,N_t)
\]

subject to

\[
\tau_{q,t}=\tau_{q,c}+d
\]

or equivalently

\[
L(\tau_{q,c}; y_{c,(i)}<\tau_{q,c}<y_{c,(i+1)}, y_{t,(j)}-d<\tau_{q,c}<y_{t,(j+1)}-d)=h(i|q,N_c)\times h(j|q,N_t)
\]

The constrained maximization may be performed as a simple line search over $\tau_{q,c}$ between the unconstrained maximizing values (i.e. sample quantiles possibly shifted by $d$).

Suppose the constrained joint likelihood is maximized at order statistics $i^*_d$ and $j^*_d$. Then a likelihood ratio test statistic can be formed as

\[
LR=-2\ln\left[\frac{h(i^*_d|q,N_c)\times h(j^*_d|q,N_t)}{h(\lfloor q(N_c+1)\rfloor |q, N_c) \times h(\lfloor q(N_t+1)\rfloor|q, N_t)}\right]
\]

Denote this quantity $H(i^*_d,j^*_d|q,N_c,N_t)$. By Wilks' theorem,

\[
\lim_{N_c\to \infty,N_t\to \infty} H(i^*_d,j^*_d|q,N_c,N_t) \sim \chi^2(1)
\]

and thus we have the basis for a hypothesis test and therefore an acceptance region (confidence interval) on $d$:

\[
\lim_{N_c\to \infty,N_t\to \infty} C(\alpha)=\{d : H(i^*_d,j^*_d|q,N_c,N_t) < \chi^2_\alpha(1) \}
\]

\begin{figure}[h]
\centering
\includegraphics[width=\textwidth]{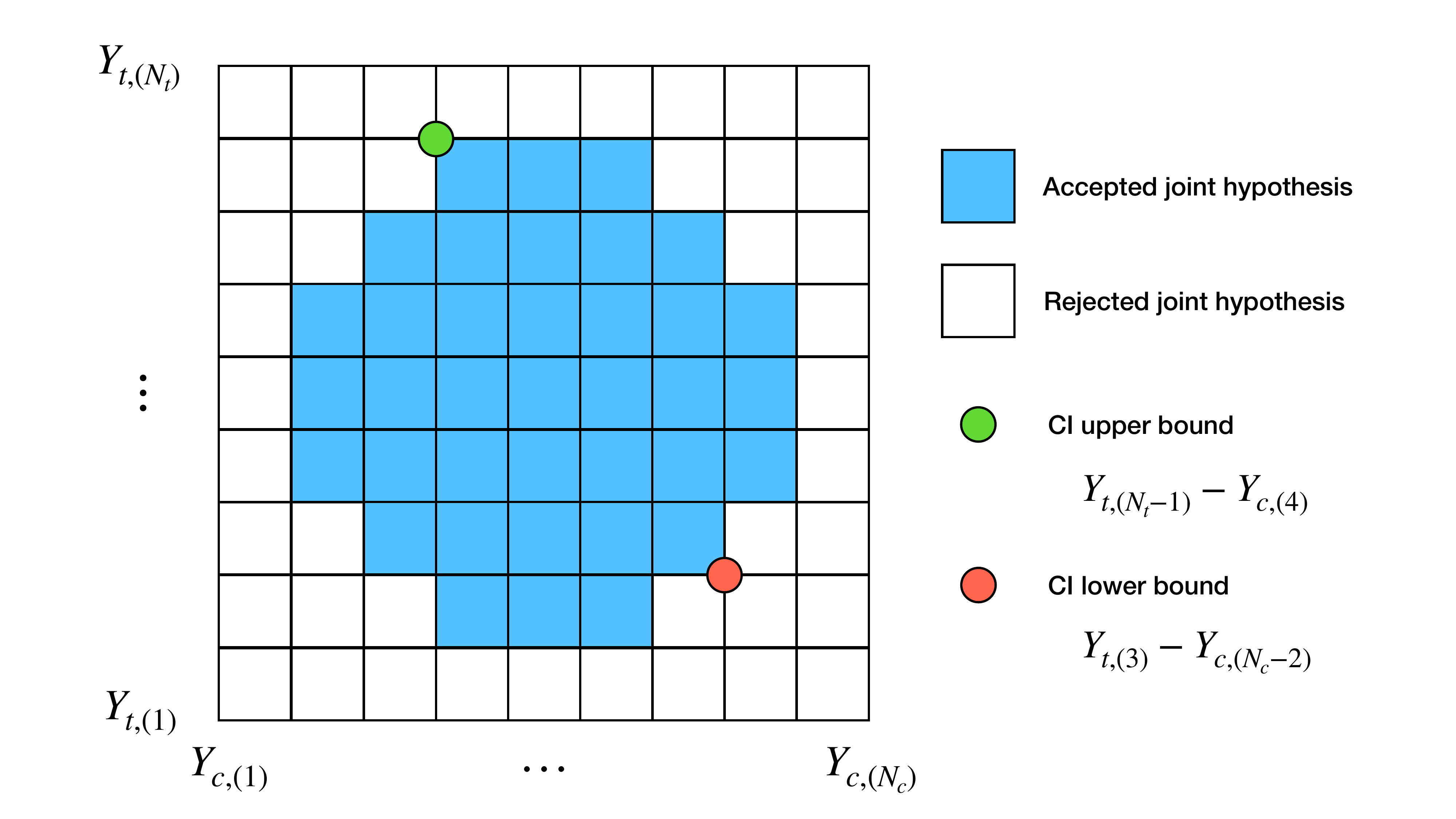}
\caption{Visualization of the difference-in-quantile acceptance region}
\end{figure}

The problem may be inverted to simplify computation. Search the acceptance region for extreme values of $d$ as

\[
C(\alpha)=(\min_{i,j}y_{t,(j)}-y_{c,(i)}, \max_{i,j} y_{t,(j)}-y_{c,(i)}) \quad {\rm s.t.} \quad H(i,j|q,N_c,N_t) < \chi^2_\alpha(1)
\]

The acceptance region can be visualized as an $N_c \times N_t$ grid of points; the ``tiles'' between points represent areas where the joint likelihood is constant. Each cross-sample pair of points represents a potential extreme value in the acceptance region.

Simulations show that a search of the acceptance region for extreme values yields a conservative confidence interval; the conservativeness is likely due to multiple comparisons. In the next section, structure will be added to the problem to permit a single hypothesis test that produces a confidence interval closer to the nominal coverage.

\section{Large sample behavior}

By the de Moivre--Laplace theorem, a binomial converges to a normal distribution and

\[
\lim_{N\to\infty} h(i|q,N)=\frac{1}{\sqrt{2\pi Nq(1-q)}}\exp\left(-\frac{1}{2}\frac{(i-Nq)^2}{Nq(1-q)}\right)
\]

Thus

\[
\lim_{N_c\to\infty,N_t\to\infty} H(i,j|q,N_c,N_t) = \frac{(i-N_c q)^2}{N_c q(1-q)}+\frac{(j-N_t q)^2}{N_t q(1-q)}
\]

and so

\[
C(\alpha)=(\min_{i,j}y_{t,(j)}-y_{c,(i)}, \max_{i,j} y_{t,(j)}-y_{c,(i)}) \quad {\rm s.t.} \quad \frac{(i-N_c q)^2}{N_c q(1-q)}+\frac{(j-N_t q)^2}{N_t q(1-q)}
 < \chi^2_\alpha(1)
\]

To avoid a complete search of the ellipse perimeter, suppose that the true cdfs $F_c$ and $F_t$ are each linear inside the elliptical ball:

\[
\begin{array}{c}
F_c(y) = m_c y + b_c \\
F_t(y) = m_t y + b_t
\end{array}
\]

Then in this region,

\[
\lim_{N_c\to\infty,N_t\to\infty}y_{t,(j)}-y_{c,(i)}=F_t^{-1}(j/N_t) - F_c^{-1}(i/N_c)=\left(\frac{j}{m_t N_t} - b_t\right)
 - \left(\frac{i}{m_c N_c} - b_c\right)
 \]

A confidence interval can then be found by maximizing and minimizing this expression with respect to $i$ and $j$ subject to the elliptical $\chi^2$ constraint. Form the Lagrangian

\[
\mathcal{L}=\left(\frac{j}{m_t N_t} - b_t\right) - \left(\frac{i}{m_c N_c} - b_c\right)+\lambda \left[ \frac{(i-N_cq)^2}{N_cq(1-q)} + \frac{(j-N_tq)^2}{N_tq(1-q)} - \chi^2_\alpha(1) \right]
\]

The first-order conditions are

\[
\frac{\partial \mathcal{L}}{\partial i}=-\frac{1}{N_cm_c}+\lambda\frac{2(i-N_cq)}{N_cq(1-q)}=0
\]

\[
\frac{\partial \mathcal{L}}{\partial j}=-\frac{1}{N_tm_t}+\lambda\frac{2(j-N_tq)}{N_tq(1-q)}=0
\]

\[
\frac{\partial \mathcal{L}}{\partial \lambda}=
\frac{(i-N_cq)^2}{N_cq(1-q)} + \frac{(j-N_tq)^2}{N_tq(1-q)} - \chi^2_\alpha(1)=0
\]

Solving, we find the optimal indexes to be

\[
i=N_cq \pm z_{\alpha/2}\sqrt{\frac{N_cN_tq(1-q)}{N_t+N_cm_c^2/m_t^2}}
\]

\[
j=N_tq \pm z_{\alpha/2}\sqrt{\frac{N_cN_tq(1-q)}{N_c+N_tm_t^2/m_c^2}}
\]

And so a confidence interval on $\delta=\tau_{q,t}-\tau_{q,c}$ is given by

\[
C(\alpha)=(y_{t,(j^-)}-y_{c,(i^+)}, y_{t,(j^+)}-y_{c,(i^-)})
\]

Implementing this method requires knowledge of $m_c$ and $m_t$. If the treatment effect is constant within the elliptical acceptance region, these can be assumed equal and will drop out of the index equations. But if the treatment effect is (linearly) heterogeneous within the acceptance region, it will be necessary to estimate $m_c$ and $m_t$ from the data.

The following two-step algorithm can be used to form a final confidence interval:

\begin{enumerate}
\item Compute $i^-_1$, $i^+_1$, $j^-_1$, and $j^+_1$ under the assumption that $m_c=m_t$ as
\[
\begin{array}{c}
i_1=N_cq \pm z_{\alpha/2}s \\
j_1=N_tq \pm z_{\alpha/2}s
\end{array}
\]
where
\[
s=\sqrt{\frac{N_cN_tq(1-q)}{N_c+N_t}}
\]
\item Estimate
\[
\hat{m}_c=\frac{F_c(y_{c,(i^+_1)})-F_c(y_{c,(i^-_1)})}{y_{c,(i^+_1)}-y_{c,(i^-_1)}}=\frac{(i^+_1-i^-_1)/N_c}{y_{c,(i^+_1)}-y_{c,(i^-_1)}}
\]
and
\[
\hat{m}_t=\frac{F_t(y_{t,(j^+_1)})-F_t(y_{t,(j^-_1)})}{y_{t,(j^+_1)}-y_{t,(j^-_1)}}=\frac{(j^+_1-j^-_1)/N_t}{y_{t,(j^+_1)}-y_{t,(j^-_1)}}
\]
\item Compute $i^-_2$, $i^+_2$, $j^-_2$, and $j^+_2$ using the estimated $\hat{m}_c$ and $\hat{m}_t$
\end{enumerate}

In this way, a confidence interval is formed using only four points from each ordered sample, and without a perimeter or region search. Simulations indicate that this confidence interval is less conservative than a full ellipse search, and quite close to the nominal coverage level.

\bibliographystyle{plain}
\bibliography{quantiles}

\begin{thebibliography}{1}

\bibitem{zou2008}
A.~Donner G.~Y.~Zou.
\newblock Construction of confidence limits about effect measures: A general
  approach.
\newblock {\em Statistics in Medicine}, 27(10):1693--702, 2008.

\bibitem{price2002}
Robert~M. Price and Douglas~G. Bonett.
\newblock Distribution-free confidence intervals for difference and ratio of
  medians.
\newblock {\em Journal of Statistical Computation and Simulation},
  72(2):119--124, 2002.

\bibitem{spotifybootstrap}
M{\aa}rten Schultzberg and Sebastian Ankargren.
\newblock Resampling-free bootstrap inference for quantiles.
\newblock In Kohei Arai, editor, {\em Proceedings of the Future Technologies
  Conference (FTC) 2022, Volume 1}, pages 548--562, Cham, 2023. Springer
  International Publishing.

\end{thebibliography}

\end{document}